\documentclass[fleqn,12pt,twoside]{article}
\usepackage{espcrc1}

\usepackage{graphicx}

\newcommand{\AmS}{{\protect\the\textfont2
  A\kern-.1667em\lower.5ex\hbox{M}\kern-.125emS}}

\title{Study of the ${^3\mbox{He}}-\eta$ system in $d-p$ collisions}

\author{
J.~Smyrski\address[Cracow]{Institute of Physics, Jagiellonian University, 
Pl-30-059 Cracow, Poland} \thanks{Corresponding author, e-mail: smyrski@if.uj.edu.pl},
H.-H.~Adam\address[Munster]{Institut f\"ur Kernphysik, Westf\"alische 
Wilhelms-Universit\"at, D-48149 M\"unster, Germany},
A.~Budzanowski\address[Bronowice]{Institute of Nuclear Physics, Pl-31-342 Cracow, Poland},
E.~Czerwi\'nski\addressmark[Cracow],
R.~Czy\.zykiewicz\addressmark[Cracow],
D.~Gil\addressmark[Cracow],
D.~Grzonka\address[Julich1]{IKP, ZEL, Forschungszentrum J\"ulich, 
D-52425 J\"ulich, Germany},
M.~Janusz\addressmark[Cracow],
L.~Jarczyk\addressmark[Cracow],
B.~Kamys\addressmark[Cracow],
A.~Khoukaz\addressmark[Munster],
P.~Klaja\addressmark[Cracow],
P.~Moskal\addressmark[Cracow],
W.~Oelert\addressmark[Julich1],
C.~Piskor-Ignatowicz\addressmark[Cracow],
J.~Przerwa\addressmark[Cracow],
J.~Ritman\addressmark[Julich1],
T.~Ro\.zek\address[Katowice]{Institute of Physics, University of Silesia, 
PL-40-007 Katowice, Poland},
T.~Sefzick\addressmark[Julich1],
M.~Siemaszko\addressmark[Katowice],
A.~T\"aschner\addressmark[Munster],
P.~Winter\addressmark[Julich1],
M.~Wolke\addressmark[Julich1],
P.~W\"ustner\addressmark[Julich1]
and~W.~Zipper\addressmark[Katowice]
}
    
\begin{document}

\maketitle

\begin{abstract}
We have measured excitation functions for the 
$dp \rightarrow {^3\mbox{He}}\,X,\, (X=\pi^0, \eta)$ channels 
near the $\eta$ production threshold.
The data were taken during a slow ramping of the COSY internal deuteron beam
scattered on a proton target.
The excitation function for the reaction $dp \rightarrow {^3\mbox{He}}\,\pi^0$
does not show any structure which could originate from the decay 
of ${^3\mbox{He}}-\eta$ bound state.
We measured also the threshold excitation curve 
for the $dp \rightarrow {^3\mbox{He}}\,X$ process,
however, contrary to the SATURNE results, we observe no cusp near the $\eta$ threshold.
\end{abstract}

\section{INTRODUCTION}

Study of the interaction between the $\eta$ meson and the ${^3\mbox{He}}$ nucleus
is of high interest due to the possible existence of a bound state \cite{Wilk93,Pfei04}
and because of the potentiality of a strict description of this system within
the four body scattering theory \cite{Aren02}.
The $d-p$ collision is very well suited for the creation of 
a ${^3\mbox{He}-\eta}$ pair since the corresponding production cross section 
(about 0.4~$\mu$b) is relatively high close to threshold.
Investigations can be conducted both: above threshold, where the absolute value
of the scattering length can be determined on the basis of FSI effects \cite{Sibi04}, and below 
threshold, where one should search for resonance-like structures 
in the excitation functions of decays from the eventual ${^3\mbox{He}} - \eta$ bound 
state in various possible reaction channels like 
e.g. in the ${^3\mbox{He}}\,\pi^0$ channel \cite{Baru03}.
In order to study these topics we performed measurements of the exciatation
functions for the $dp \rightarrow {^3\mbox{He}}\,X,\, (X=\pi^0, \eta)$ reactions 
near the $\eta$ production threshold.

\section{EXPERIMENT}

\begin{figure}[htb]
\begin{minipage}[t]{70mm}
\includegraphics[height=2.in]{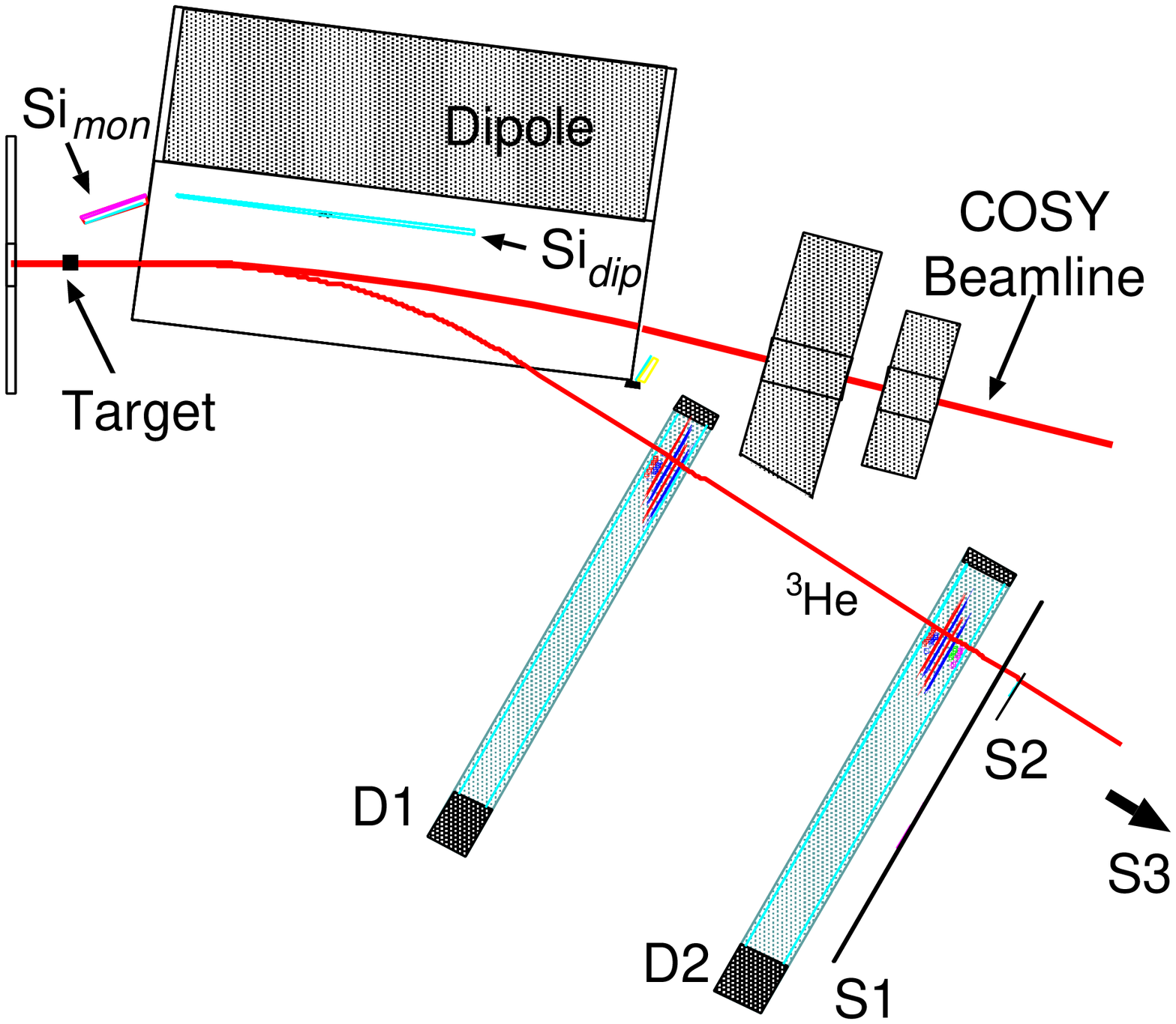}
\caption{Scheme of the COSY-11 detection system.}
\label{fig:c11}
\end{minipage}
\hspace{\fill}
\begin{minipage}[t]{70mm}
\includegraphics[height=2.in]{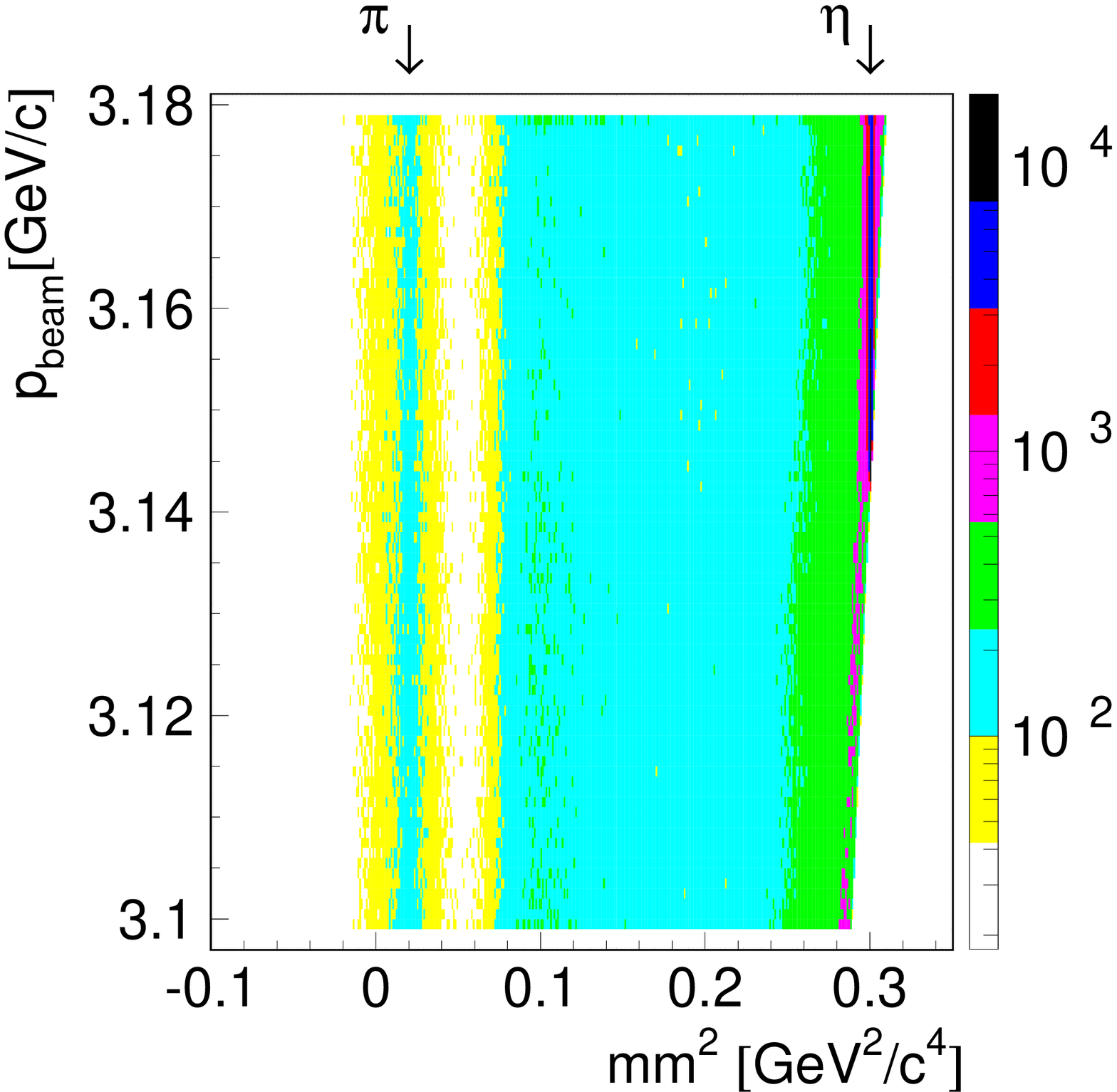} 
\caption{Missing mass (x-axis) as a function of beam momentum (y-axis).
}
\label{fig:mm}
\end{minipage}
\end{figure}

The experiment was performed at the COoler SYnchrotron COSY in J\"ulich
with the use of the COSY-11 detection facility \cite{Brau96} shown schematically in Fig.~\ref{fig:c11}.
The internal deuteron beam of COSY was scattered on a hydrogen cluster target installed in front of a COSY accelerator dipole magnet.
The outgoing ${^3\mbox{He}}$-ions from the $dp \rightarrow {^3\mbox{He}}\,X$ reactions 
were momentum analyzed in the dipole magnet and their trajectories were
registered with two drift chambers D1 and D2.
Identification of the ${^3\mbox{He}}$-ions was based on the energy loss 
in the scintillation hodoscope S1 and, independently,
on the time of flight on a path of 9~m between the scintillation hodoscope S1 and S3.

\begin{figure}[htb]
\begin{minipage}[t]{75mm}
\includegraphics[width=2.2in]{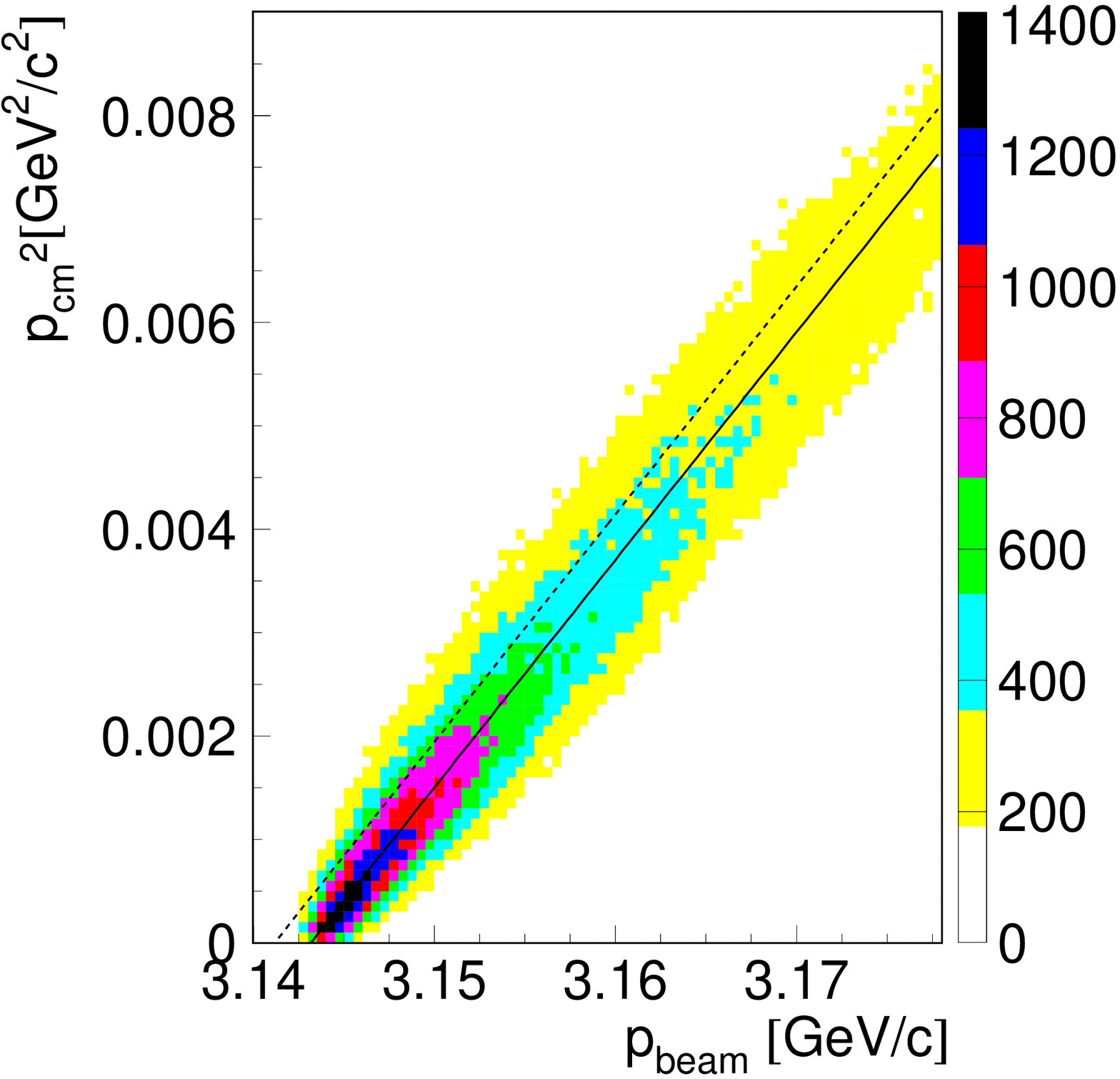}
\caption{${^3\mbox{He}}$ c.m. momentum squared as a function of the nominal beam momentum.
Solid line corresponds to a fit to the experimental counts and dashed line represents
kinematical relation based on known particle masses.}
\label{fig:ecor}
\end{minipage}
\hspace{\fill}
\begin{minipage}[t]{75mm}
\includegraphics[width=2.2in]{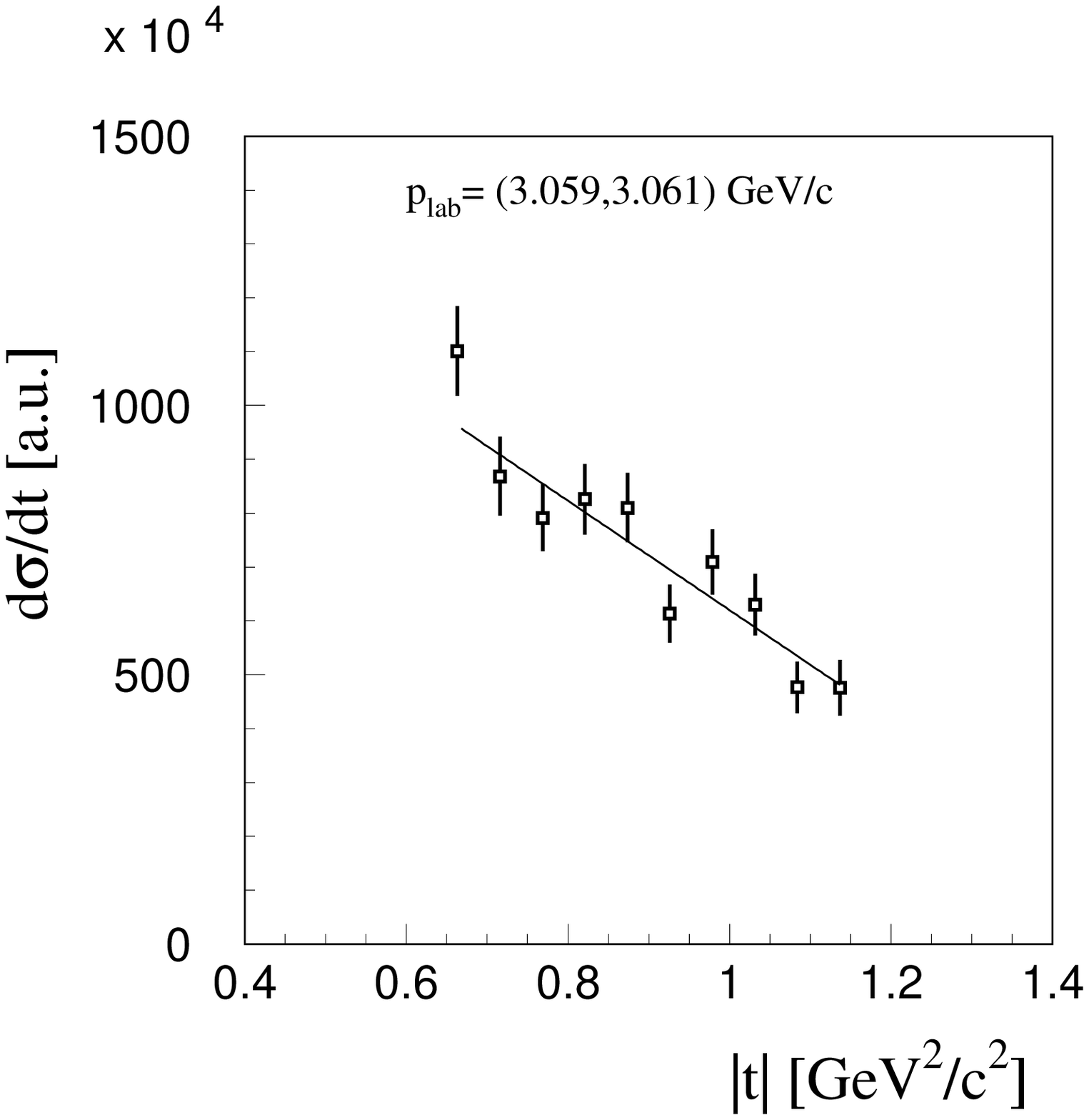} 
\caption{Experimental rate of elastic $dp$ events corrected for the detector acceptance 
as a function of the four-momentum transfer. Solid line corresponds to parametrization
of elastic $dp$ cross sections.}
\label{fig:lumidp}
\end{minipage}
\end{figure}

The momentum of the deuteron beam was varied continuously within each cycle from 3.095~GeV/c
to 3.180~GeV/c, crossing the threshold for the $dp \rightarrow  {^3\mbox{He}}\,\eta$
reaction at 3.141~GeV/c.
In the missing mass spectrum determined as a function of the beam momentum 
(see Fig.~\ref{fig:mm}) 
a clear signals from the $\eta$ meson production as well as from the single $\pi^0$ production are visible.
The present nominal beam momenta in the range around 3.1~GeV/c,
calculated from the synchrotron frequency and the beam orbit length,
are know at COSY with accuracy of 3~MeV/c. 
We determined the beam momentum more precisely using the dependence 
of the ${^3\mbox{He}}$ c.m. momentum squared ($p_{cm}^2$)
on of the beam momentum for the $dp \rightarrow {^3\mbox{He}} \eta$ reaction.
The solid line in Fig.~\ref{fig:ecor} represents a fit 
to the experimental points and the dashed line represents
relation based on the two-body kinematics.
The threshold beam momenta determined for these two cases,
corresponding to $p_{cm} = 0$~MeV/c, differ by $\Delta p = (-2.0 \pm 0.4$~MeV/c. 
This difference was taken as a correction to the nominal beam momentum.
The indicated error is dominated by contribution due to the uncertainty 
of the mass of the $\eta$ meson ($547.75 \pm 0.12$)~MeV \cite{Revi04} 
which influences the present result via the threshold energy.

The luminosity was determined by comparing differential counting rates
for simultaneously registered elastic $d-p$ scattering with corresponding
cross sections taken from literature and parametrized as a function of 
the four-momentum transfer (see Fig.~\ref{fig:lumidp}).

\section{EXCITATION FUNCTIONS}

\begin{figure}[htb]
\begin{minipage}[t]{75mm}
\includegraphics[width=3.in]{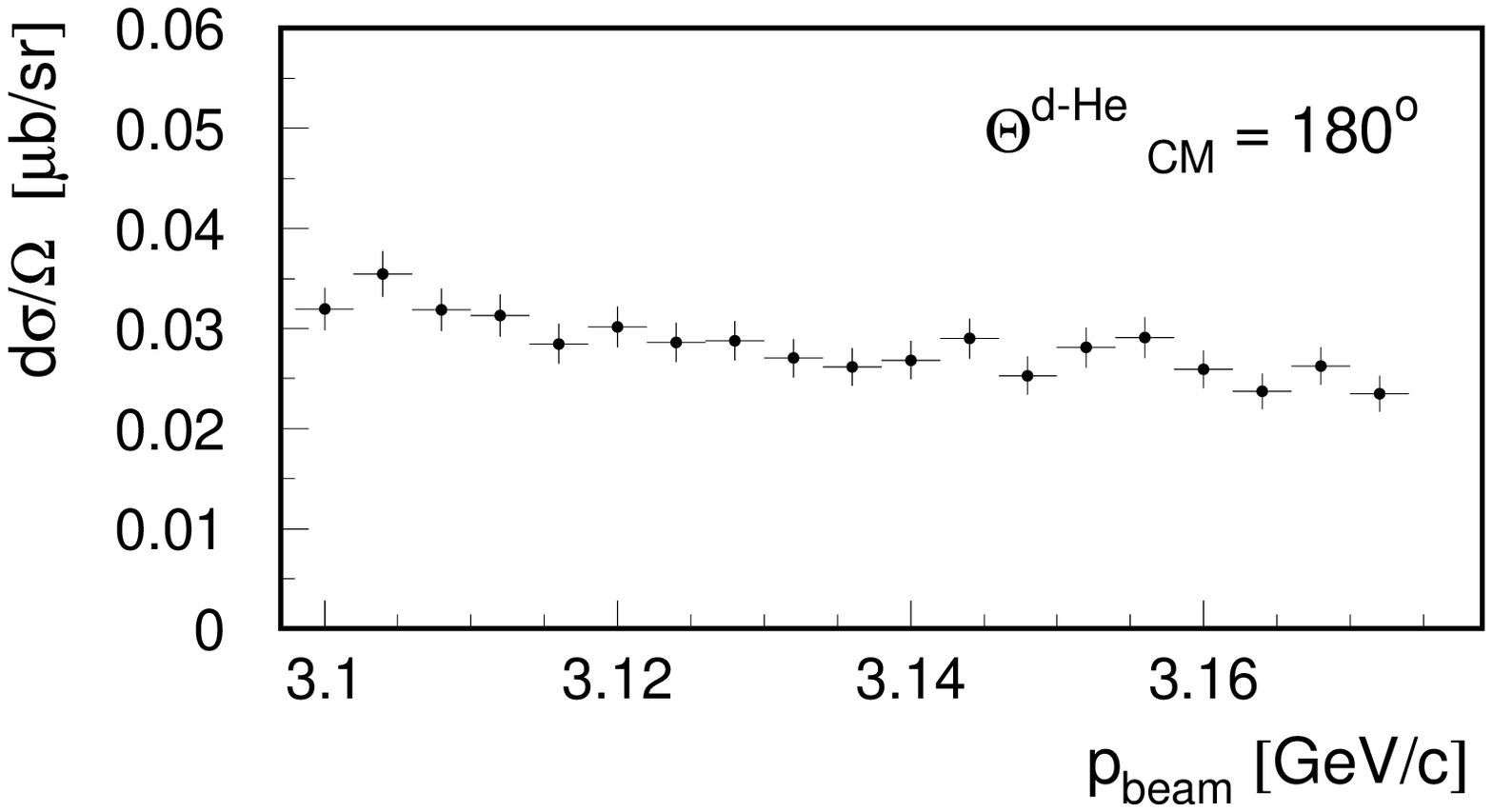}
\caption{$dp \rightarrow {^3\mbox{He}}\,\pi^0$ cross section for backward ${^3\mbox{He}}$ scattering as a function of beam momentum.}
\label{fig:pion}
\end{minipage}
\hspace{\fill}
\begin{minipage}[t]{70mm}
\includegraphics[width=2.in]{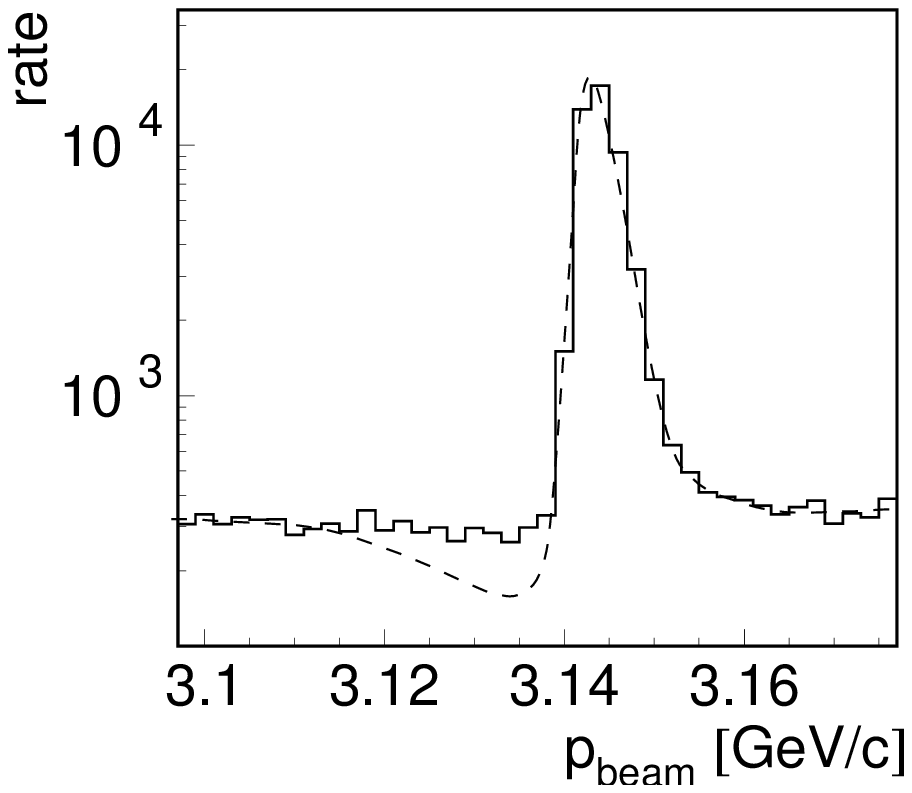} 
\caption{Threshold excitation curve for the $dp \rightarrow {^3\mbox{He}}\, X$ reaction
measured near the $\eta$ threshold.
Dashed line indicates shape of expected cusp.
}
\label{fig:excite}
\end{minipage}
\end{figure}

In our search of the ${^3\mbox{He}}-\eta$ 
bound state, we investigated the $dp \rightarrow {^3\mbox{He}}\, \pi^0$ 
differential cross sections for the forward pion angles  
($\Theta_{d-\pi}^{cm} = 0^{\circ}$). 
This choice is dictated by the fact that the $dp \rightarrow {^3\mbox{He}} \pi^0$
cross section is up to two orders of magnitude smaller at the forward angles
than at the most backward angles \cite{Silv85}.
Assuming that the searched structure is produced isotropically,
one can expect that it can be best seen just at the forward angles since
it appears on the level of small ``non-resonant'' cross section.
Fig.~\ref{fig:pion} shows the pion production cross sections 
as a function of the beam momentum.
Except of statistical fluctuations no structure can be seen in this curve.
Assuming, that a $3\sigma$ deviation in this curve would be a signal
of the ${^3\mbox{He}}-\eta$ bound state formation and its subsequent 
decay in ${^3\mbox{He}} \pi^0$ channel, we determined the upper limit 
for the corresponding cross section as equal to 70~nb. 
This limit appears not very restrictive at least under assumption
that the cross sections for the ${^3\mbox{He}}-\eta$ bound state formation are of the same
order as the $dp \rightarrow {^3\mbox{He}}\, \eta$ cross sections near threshold (0.4$\mu$b), 
and that other possible decay channels like  $dp \pi^0$ are more favorable.

The analysis of the experimental data for the $dp \rightarrow {^3\mbox{He}}\, \eta$ channel
is not finished yet, however, the preliminary total cross sections for excess energies $Q < 4$~MeV are in agreement with the data
of Mayer et al.~\cite{Maye96} and confirm the strong enhancement observed near-threshold 
due to the ${^3\mbox{He}}-\eta$ FSI.

Data collected in the present experiment 
were also used to investigate the cusp observed in the threshold excitation curve for 
the $dp \rightarrow {^3\mbox{He}}\, X$ process which was measured 
with the SPES-IV spectrometer at SATURNE~\cite{Plou88}. 
The cusp was visible at the $\eta$ threshold and,
as suggested by Wilkin \cite{Wilk88}, it can be caused by an interference between an intermediate state including
the $\eta$ meson and the non-resonant background corresponding to the multi-pion production.
The threshold excitation curve was determined by varying
the beam momentum and adjusting the setting of the SPES-IV spectrometer in such a way 
that only the ${^3\mbox{He}}$ associated with the maximum missing masses were registered. 
Since the COSY-11 momentum and angular acceptance is much larger than one of the SPES~IV spectrometer, limitation of the acceptance was realized by means of corresponding cuts during the data analysis. More details concerning the data analysis and preliminary threshold excitation curve are given in Ref.~\cite{Smyr06}.
Final threshold excitation curve based on the full collected statistics
is shown in Fig.~\ref{fig:excite}.
The peak in the middle of the spectrum is associated with opening 
of the $dp \rightarrow {^3\mbox{He}}\, \eta$ channel. 
Contrary to the SATURNE result we see no cusp near the $\eta$ threshold.

%

\end{document}